\documentclass[twocolumn, aps]{revtex4}
\usepackage{times}
\usepackage{graphics}

\begin{document}

\title{Simulations of Oligomeric Intermediates in Prion Diseases}
\author{David L. Mobley\footnote{To whom reprint requests should be addressed, Tel.: 530-752-0446; Fax: 530-752-4717; E-mail: mobley@physics.ucdavis.edu}}
\affiliation{Department of Physics, University of California, Davis, CA 95616}
\author{Daniel L. Cox}
\affiliation{Department of Physics, University of California, Davis, CA 95616}
\author{Rajiv R. P. Singh}
\affiliation{Department of Physics, University of California, Davis, CA 95616}
\author{Rahul V. Kulkarni\footnote{Current address NEC Research Labs, 4 Independence Way, Princeton, N.J. 08540}}
\affiliation{Department of Physics, University of California, Davis, CA 95616}
\author {Alexander Slepoy}
\affiliation{MS 0316, Sandia National Laboratories, Albuquerque, NM 87175-0316}

\date{\today}

\begin{abstract}
We extend our previous stochastic cellular automata based model for areal aggregation of prion proteins on neuronal surfaces.  The 
new anisotropic model allow us to simulate both strong $\beta$-sheet and weaker attachment bonds between proteins. Constraining binding directions allows us to generate aggregate structures with 
the hexagonal lattice symmetry found in recently observed \emph{in vitro} experiments.  We argue that these constraints on rules may correspond to underlying steric constraints on the aggregation process.  We find that 
monomer dominated growth of the areal aggregate is too slow to account for some observed doubling time-to-incubation time ratios inferred from data, 
and so consider aggregation dominated by relatively stable but non-infectious oligomeric intermediates.  We compare a kinetic theory analysis of oligomeric 
aggregation to spatially explicit simulations of the process.  We find that with suitable rules for misfolding of oligomers, possibly due to water exclusion by the surrounding aggregate, the resulting oligomeric aggregation model maps onto our previous monomer aggregation model. Therefore it can produce some of the same attractive features for the description of prion incubation time data. We propose experiments to test the oligomeric aggregation model.

\end{abstract}

\maketitle

\section{INTRODUCTION}

Prion diseases are a group of neurodegenerative diseases including
bovine spongiform encephalopathy (BSE) in cattle, scrapie in
sheep and goats, chronic wasting disease in deer and
elk, and kuru and Creutzfeldt-Jakob Disease (CJD) in
humans. These diseases came to the forefront after BSE reached
epidemic proportions in Great Britain in the early 1990s, and
it was later shown that transmission of BSE to humans can lead
to new variant CJD (vCJD) in humans (Bruce et al., 1997),
(Hill et al., 1997), (Scott et al., 1999).

Prion diseases are unusual in that they appear to be caused by
infection with some minimal infectious "seed" of misfolded prion
protein, which alone may be able to cause disease by catalyzing
further misfolding and, in many cases, aggregation of the prion
protein. These aggregates are typically amyloid-like fibrils or
amyloid plaques (Caughey, 2000). The infectious agent is unusually
hard to eliminate by various methods including UV irradiation, suggesting it contains no nucleic acid and ratheronly protein, the so-called "protein-only" hypothesis in prion
diseases(Weissmann et al., 2002). 

In the case of CJD, a sporadic form of the diseases also exists,
occurring more or less randomly worldwide with an incidence of about
one in a million people per year. It has been suggested that this
incidence is due to the very rare event of nucleating the minimal
infectious seed by chance in a healthy individual (Come et al., 1993).
 
Developing an understanding of these diseases is important because,
for one, they are invariably fatal. To date, no treatment
exists. Additionally, it is not yet clear how large the vCJD epidemic
in humans will be; an understanding of the disease process is
important to be able to guide the search for treatment ideas.

In many cases, prion diseases result in large, up to micron-scale
plaques in the brains of people and animals with these diseases. They
also involve vacuolization or spongiform change in the brain due to
death of neurons(Scott et al., 1996). Additionally, the normal form of
the prion protein (known as PrP\(^C\)) has long been known to misfold
and aggregate \emph{in vitro} when catalyzed by the presence of a misfolded
prion protein (PrP\(^{Sc}\)) seed (Come et al., 1993). Together, these
observations have suggested to some that the aggregation process
itself may be important in these diseases (Come et al., 1993; Masel
et al., 1999). It has also been suggested that the rate-limiting step
in aggregation is nucleation of an appropriate seed, thus the rapid
aggregation in the seeded case described above (Come et al., 1993).

Another fact which may be important to this issue is that the prion
protein is normally GPI-anchored to the cell surface. Aggregation \emph{ in
vitro} as mentioned above is observed in solution rather than in the
presence of the GPI anchor on a cell surface, leaving the possibility
that the aggregation process in vivo is different.

Aggregation models developed to explore the aggregation process in
prion disease include one-dimensional, fibrillar
aggregation-and-fission models(Masel et al., 1999; Slepoy et al.,
2001), since aggregates grown \emph{in vitro} are typically seen to be
fibrillar. Additionally, our earlier work suggested that a
two-dimensional areal aggregation model could explain certain other
properties of the diseases (Slepoy et al., 2001). This earlier model
is attractive in that it can provide a simple explanation for the long
lag phase which is sometimes observed in growth of the amount of
infectious material in the brain. This lag phase of little or no
growth is followed by a doubling phase with a short characteristic
doubling time. Additionally, our earlier model provides a possible
explanation of some of the difference between infectious and sporadic
forms of CJD (Slepoy et al., 2001). In later work, we used this model
to explain and fit experimental dose incubation curves (Kulkarni et
al., 2002).

However, there were drawbacks to the earlier aggregation model we
proposed. First, no such areal aggregates had so far been
observed. Second, the fissioning essential to the model would involve
breaking of strong bonds between the proteins, probably bonds between
$\beta$-sheets (Serag et al., 2002).

More recent experimental work found two-dimensional areal aggregates of
prion protein produced during the purification process. These aggregates were examined under electron microscope (EM) and found to consist trimeric or hexameric subunits. These subunits are linked together in a regular array, possibly by their N-terminal sugars or a weak protein-protein interaction (Wille et al., 2002).

This suggested we should modify our earlier model and attempt to
reproduce this aggregate morphology. We thought of two basic schemes
for growing aggregates of this sort: (1) Growing the aggregate outward
monomer by monomer from an initial seed, or (2) Oligomeric
intermediates (possibly very flexible and of
unstable shape), form on their own in solution and are only catalyzed
into stably misfolding in the presence of an existing misfolded
seed. Some evidence in favor of (2) has already been produced:
Monomers of yeast prion can form intermediates if left to stand, which
allows aggregation to proceed at an initial faster rate when catalyzed
by addition of a seed (Serio et al., 2000). Additionally, the conformation-dependent immunoassay developed by Safar et al. (2002) detects both protease-sensitive and protease-resistant PrP\(^{Sc}\). In hamster brains, sensitive PrP\(^{Sc}\) is observed earlier, followed by resistant PrP\(^{Sc}\). This could correspond to (2) above, where the sensitive PrP\(^{Sc}\) is the intermediates that are not yet stably misfolded and the resistant PrP\(^{Sc}\) is stably misfolded intermediates. 


Work here has been done to further explore these two potential
modifications of our earlier model to examine whether they retain the
same features and if additional insight can be gained.

It is important to note that even if areal aggregation is not
important to the time course of these diseases, the aggregates
observed by Wille et al. have already provided insight into the
structure of the misfolded prion protein (Wille et al.,
2002). Theoretical modeling may be able to place further constraints
on the protein or subunit structure necessary to reproduce these
aggregates, and hence provide valuable information because these
aggregates \emph{can} form, even if they are not important to the disease
progression.

\section{BASICS OF OUR MODEL}

Here we explore the two basic schemes suggested above for growing
aggregates like those observed by Wille et. al. To do so, we use a
modification of our earlier model. Therefore a recap of common
features of these models is useful.

These models are stochastic cellular automata models, meaning that they take
place on a lattice with probabilistic interaction and diffusion rules governing
 the progression of the system. In this case, sites on the lattice are
either occupied by individual prion proteins, or water (empty, in the
simulation). The protein form at a site can also vary from PrP\(^C\)
to PrP$^{Sc}$.

Rules vary depending on the model being explored, but the basic
procedure is the same. Every simulation step, which represents a small
amount of time, we allow proteins and any aggregates to diffuse a
small amount on the lattice (each object has a probability
$1/(size)^{1/2}$ of moving one lattice site in a given step). Then we
look at every protein in the lattice and update its state according to
the rules. For example, in our original model, the conformation of an
individual prion protein is determined solely by its number of
neighboring prion proteins, and this can vary from step to step. After
doing this, we add more normal prion monomers to replace any that
converted to PrP$^{Sc}$. This is due to the assumption that this
process would be taking place in a small area on a cell, and the
normal prion monomers would be added by the cell or diffuse in from
other locations on the cell surface to keep the monomer concentration
relatively constant.

\section{MONOMER GROWTH MODEL}

First, case (1) from above was explored. Simple rules were developed
(figure 1) which can reproduce aggregates similar to those observed by
Wille et al. It is important to note that though the rules were
designed to reproduce such aggregates, most modifications of these
rules could not do so. This means that the rules provide some
constraints on the protein-protein interactions necessary to reproduce
such aggregates. Also, for the purposes of this model, we are assuming the subunits are hexameric, but the corresponding model for trimeric intermediates is actually much simpler than this model and will produce similar results.

\begin{figure*}
\includegraphics{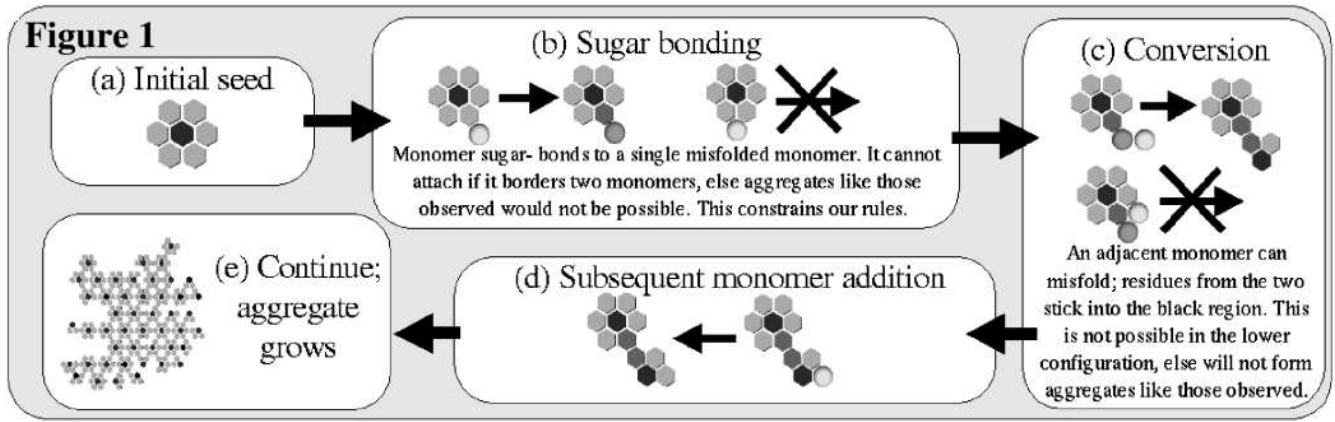}
\caption{Simple rules for monomer-by-monomer growth of aggregates like those observed. Some possible rules can be excluded, thus these rules give insight into how the proteins involved must be interacting with one another. (a) The initial seed consists of six misfolded monomers (light gray hexagons) surround a central region (black) which is occupied by some residues sticking into it from the adjacent six sites. (b) A healthy monomer (light gray sphere) can move adjacent to a misfolded one and attach via a sugar bond or other weak protein-protein interaction (proteins sugar bonded are colored dark gray). This cannot happen if the monomer moves into the site between two misfolded proteins. (c) Subsequent monomers can move next to the attached one and misfold and begin to form a new hexamer. Residues from the two stick into the black region, preventing anything else from moving there. This cannot happen if the second monomer is adjacent to the existing hexamer; this would produce irregular aggregates unlike those observed by Wille et al. (d) The forming hexamer can grow and finish via subsequent monomer addition. (e) Continue (a)-(d) for a long time and an aggregate like the one shown can form.}
\end{figure*}

Details of the algorithm for this model are covered in figure 2.

\begin{figure}
\includegraphics{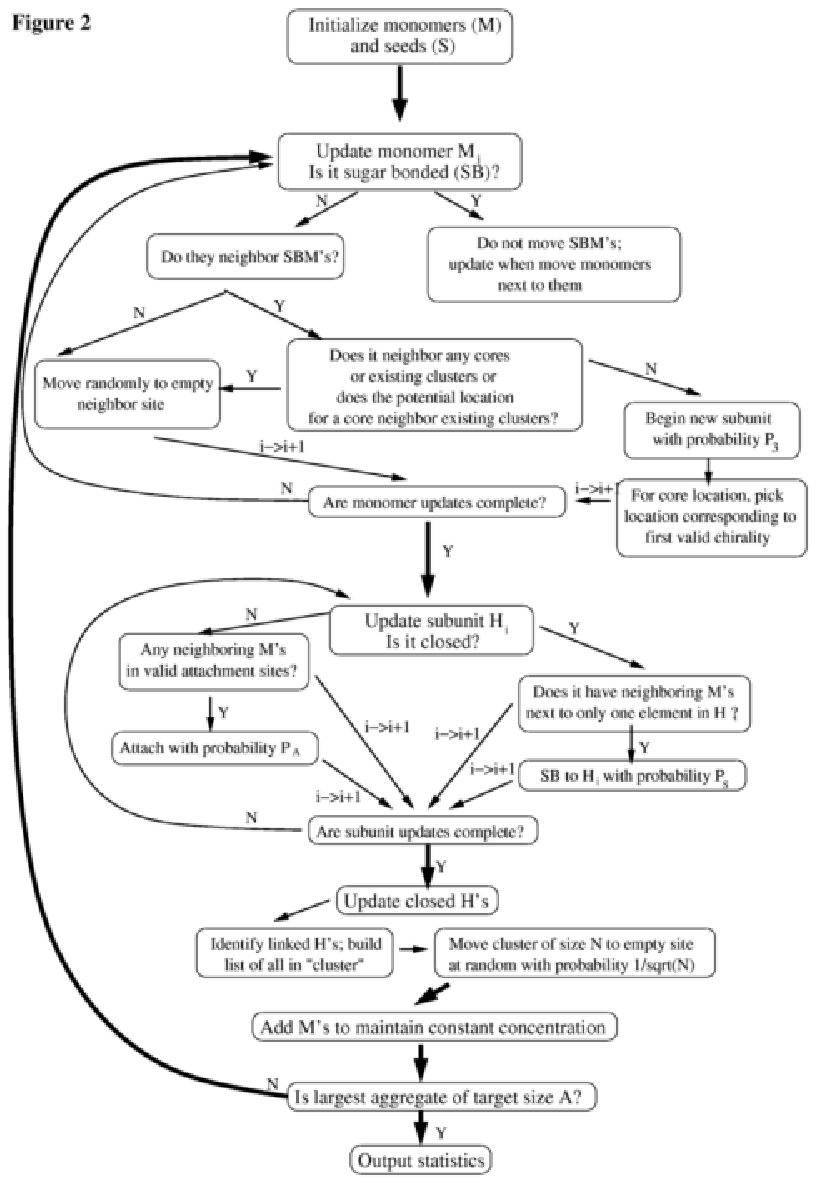}
\caption{Flow chart of simulation for monomer addition model. We typically use 
$P_3=0.2$; we tried a variety of different values for this and values near 0.2 seem to 
produce the most regular aggregates. We also typically use $P_S=0.9$. This is not 
important and roughly sets the simulation timescale. Also, for our statistics, we 
typically average over 1000 such runs as the one described here. For a larger version 
of this figure (and Fig. 6 and Fig. 10) see the author's web page at 
http://asaph.ucdavis.edu/~dmobley/} \end{figure} 

The rules are as follows: The simulation begins with a single
hexagonal subunit consisting of six misfolded monomers (light gray
hexagons in figure 1) which stick some of their residues into an
adjacent site, excluding anything else from occupying that site
(black). Healthy monomers (light gray spheres) can then attach via a
sugar-bond or other weak protein-protein interaction
to this subunit (dark gray spheres to dark gray hexagons)
but only radially outward from a monomer in the initial
hexamer. Additional monomers moving adjacent to the attached
monomer can, together with it, misfold but only if the second monomer
does not also neighbor the original hexamer. Then additional monomers
can attach to this forming hexamer, allowing it to complete. Repeating
this process many times can produce mostly regular aggregates with
some holes, similar to those observed.

The rules are also probabilistic: above, "can" means that some fraction of
the time the event occurs. These probabilities can be changed in the
simulation and give different growth rates, but the same essential
features and scaling as described below.

If this is in fact how these aggregates are forming, we find out about
the orientation of monomers within a hexagonal subunit. We find, as
mentioned in the discussion of the rules above, that the N-terminal
sugars or attachment sites 
must stick radially outward from each monomer in a hexagonalsubunit (figure 1b). This is in agreement with the hexagonal structure
proposed by Wille. Additionally, we find that no such regular
aggregates can be produced unless the monomer attaching to a
previously attached monomer (figure 1(c)) can only attach if it is not
adjacent to an existing hexamer. This seems to indicate that the other
spaces must be occupied by residues from the existing hexamer,
preventing attachment in those sites.

This model can also reproduce gaps in aggregates as observed. In this
model gaps are due to variations of the growth rate from average for
part of the aggregate, causing several parts of the aggregate to grow
apart and then rejoin after leaving a gap.

One reason for developing this model was to see if it would capture
the same features of the disease as our original model. Our original model explained the difference between the lag phase and the doubling phase by suggesting that the doubling phase is initiated when aggregates begin to fission, then regrow to a certain fissioning size and break again. Key to this explanation is our result that aggregation speeds up, so that the time for an aggregate to double in size from half its fission size to its fission size is much much less than the time for it to get from its initial size to its fissioning size.

To see if this model could produce the same separation of lag and doubling phases, we examined the aggregate growth rate as a function of size in this model (figure 3) and found it speeds up only slowly.  Na\"{\i}vely, one would expect
the growth rate to be roughly proportional to the square root of the
size, as the growth rate is proportional to the circumference of the
aggregate, which, assuming a circular aggregate, is $2 \pi r$. The
size of the aggregate is proportional to the area, $\pi r^2$, so the
radius is proportional to the square root of the size and thus the
rate proportional to the square root of the size. To a good
approximation, the growth rate observed here is well-fit by an offset
plus a term proportional to $(size)^{1/2}$, as expected.
	
\begin{figure}
\includegraphics{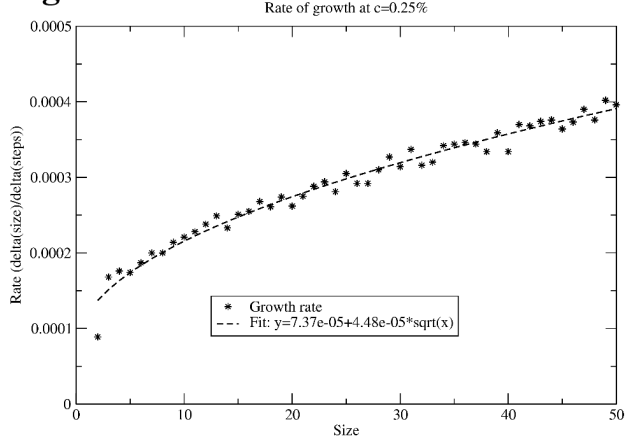}
\caption{Growth rate (change in aggregate size per step) as a function of size for seeded areal aggregation in the monomer growth model. Growth rate goes as the square root of the size with an offset, which was as expected for this model.}
\end{figure}

In this simple picture, one can calculate the ratio of the doubling
time to the lag time. The lag time is the time to go from the initial
size, say size 0 for simplicity, to size n; the doubling time from
size n/2 to size n. Integrating the rate to get the times and taking
the ratio we find $t_{doub}/t_{lag}=1-1/\sqrt{2}$ or about 0.293. This
means that this model cannot produce such a large separation between
lag and doubling times as our earlier model could, at least without
further modification.

This also indicates that if there is a lag phase and if the difference
between it and the doubling phase is due to acceleration of
aggregation, this picture is not sufficient and something more like
(2), growth from intermediates, may be a better representation of the
disease process.

\section{GROWTH VIA INTERMEDIATES}

In this case, aggregation is assumed to be the assembly of independent
hexameric intermediates into a larger areal aggregate. The
intermediates themselves are not misfolded but only misfold, in this
model, when they either aggregate with an existing misfolded seed, or
come together in such a way that they can misfold and and form a new
stable seed. In this way, the model works essentially just like the
model of Slepoy et al. except now hexameric intermediates are playing
the role of monomers (figure 4). As mentioned above, there is some
evidence that intermediates greatly increase aggregation rate in
studies of yeast prions, so this emphasis on the importance of
intermediates may be reasonable.

\begin{figure}
\includegraphics{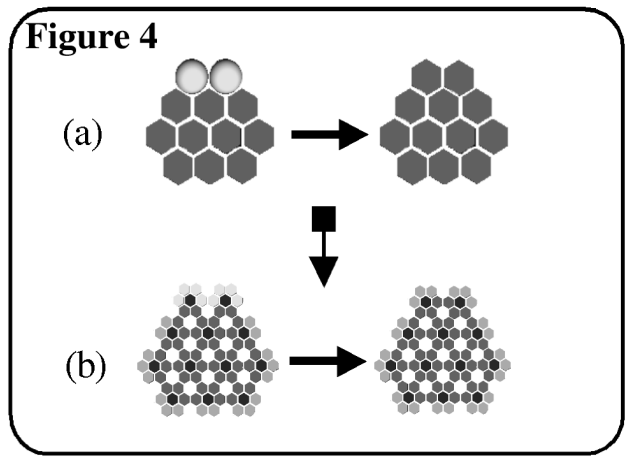}
\caption{Top, in Slepoy's model, subunits were healthy monomers (light gray spheres) aggregating with misfolded monomers (dark gray hexagons); bottom, subunits are hexagonal intermediates (light gray/dark gray) aggregating with misfolded hexagonal structures (medium gray/dark gray). In both cases, the aggregation process and kinetics ought to be, and indeed are, similar.}
\end{figure}

To be able to map this model back into our old model, though, it must
be known how the intermediate concentration depends on monomer
concentration. And this is not obvious. So a simulation was developed
to explore how the concentration of hypothetical hexameric
intermediates would depend on monomer concentration. Again, here we are assuming the intermediates are hexameric but we can easily modify the model to accomodate trimers.

To get at the concentration of intermediates, it was assumed that two monomers have a probability $P_1$
of beginning a new hexameric subunit when they come into contact (see
figure 5). This new subunit can grow by addition of monomers when they
move into appropriate positions (changing this probability does not
affect the outcome of the simulation, only the timescale, so it was
set to 1). However, this growth process competes with a "dissolving"
process by which a monomer that is part of an intermediate but only
has one neighboring monomer can break off with a probability
$P_3$. Thus the end destiny of any intermediate that begins is either
to form a complete hexameric intermediate, in which case it can
persist, or to dissolve completely.

\begin{figure}
\includegraphics{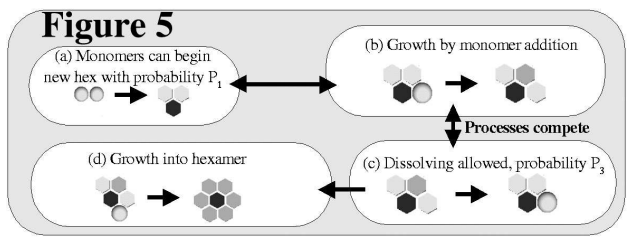}
\caption{Rules for the formation of intermediates. Note that growth and dissolving compete, so that any intermediate eventually either becomes a complete, stable hexagon or dissolves back into monomers. (a) Two monomers have a probability $P_1$ of joining to begin a new intermediate, which is not yet stably misfolded. Black represents a region blocked by some of their residues. (b) This can grow by addition of monomers to either ``end''. After attaching, the monomer sandwiched between the other two has two neighbors and is not allowed to break off, while the ones with only one neighbor can. (c) A monomer with only one neighboring monomer has a probability $P_3$ of breaking off in a given step. This competes with the growth process. (d) Continuing addition of monomers can result in a finished hexameric intermediate where every monomer has two neighbors and is safe from breaking off.}
\end{figure}

Details of the algorithm for this model are shown in figure 6.

\begin{figure}
\includegraphics{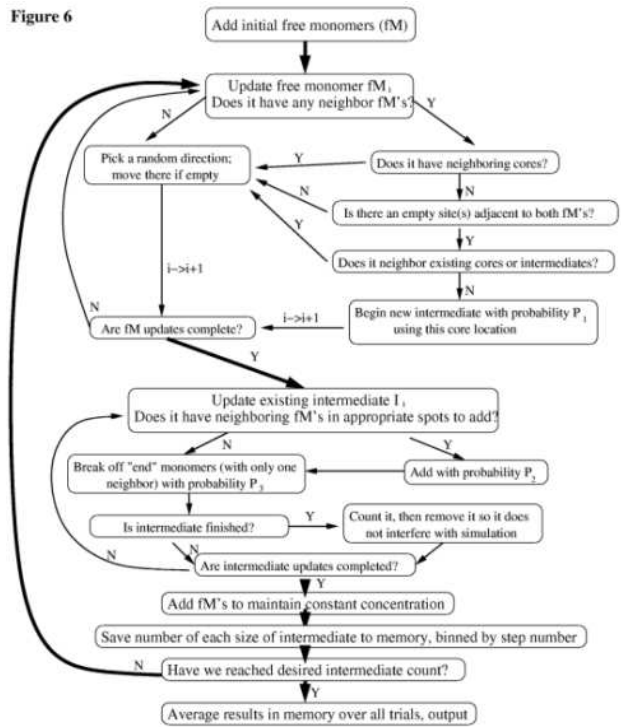}
\caption{Flow chart of simulation for the formation of intermediates. Note that $P_1$ we vary for different runs, $P_2$ we typically set to 1 (it sets the simulation timescale and is unimportant) and $P_3$ we also vary. Finished intermediates are removed so that we can run to a larger number of finished intermediates without the lattice getting clogged. Here, also, we typically average over 1000 trials for good statistics.}
\end{figure}

This dissolving, or reversibility, was included because it was not
obvious that at low monomer concentrations, one would expect a
reasonable formation rate of intermediates via this mechanism. It was
initially thought that at concentrations below something on the order
of $P_3$, breaking would dominate and the formation rate of
intermediates would be almost zero.  First, the simulation that was
developed was used to examine the dependence of time for intermediate
formation as a function of monomer concentration (figure 7). It was
found that at high monomer concentration, the time to form an
intermediate scales between $1/c$ and $1/c^2$ ($c$ is
concentration). This is because the likelihood of starting an
intermediate scales as the dimer concentration ($1/c^2$) while the
time to add monomers to it scales as $1/c$. On the other hand, at very
low monomer concentration, the time asymptotically approaches
$1/c^6$. This is due to the fact that at these concentrations,
dissolving dominates and it is only in the very rare event that six
monomers are in the same place at almost the same time that an
intermediate can finish. The probability of that scales as $1/c^6$.

\begin{figure}
\includegraphics{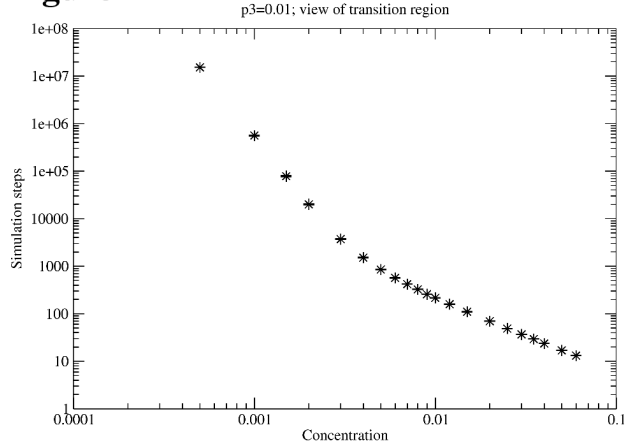}
\caption{Simulation steps (time) to form a hexameric intermediate as a function of monomer concentration. Log-log scale. Note the broad transition to dissolving-dominated behavior at low concentration. The transition actually continues to even lower concentration than can be seen here. At very low concentration the time eventually scales as $1/c^6$. Standard deviations fall within the size of the data points on this plot.}
\end{figure}

It is interesting to note that the beginning of the transition between
high concentration behavior, where most intermediates successfully
become complete, and low concentration behavior, where only a lucky
few do, begins at a concentration on the order of the breaking
probability, $P_3$. This suggests that if the strength of bonds
between intermediates could be weakened somehow, the biological number
of intermediates could be drastically decreased by pushing biological
monomer concentrations into the $1/c^6$ regime.

The goal, however, was to determine the dependence of the intermediate
concentration on monomer concentration. This just provided a formation
rate, and the functional form was uncertain. So another sort of result
was examined: We began examining behavior of the system as a function
of time, and measured the number of different partial intermediates
(two monomers, \ldots five monomers, hexameric intermediates). We
first examined the case with no breaking ($P_3 =0$) to check our
results, because it is relatively easy to work out kinetics in that
case. A sample of one of these plots is shown in figure 8, with
symbols as data points and solid lines as approximate kinetics
fits. It is important to note that in this case, and in the case of
nonzero breaking probability, the number of dimers, trimers,
tetramers, and pentamers reaches equilibrium relatively quickly and
then the hexamer number begins to grow linearly at a rate equal to the
rate of dimer formation.

\begin{figure}
\includegraphics{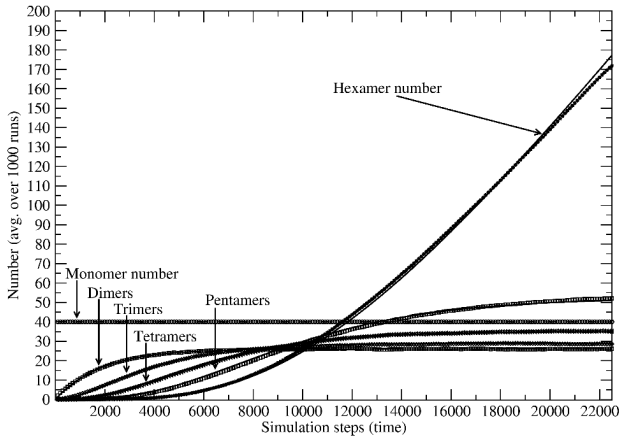}
\caption{Number of each size as a function of time (simulation steps), with zero breaking. Note at long times intermediates reach equilibrium and the hexamer number begins growing linearly with time. Points are simulation dat points; solid lines (mostly overlapping points) are approximate kinetics results.}
\end{figure}

Sample results with nonzero breaking are shown in figure 9. These
results are qualitatively similar, except the number of
pre-intermediates that persists is much lower. In the high-breaking
limit, the very low level of intermediates demonstrates that either a
potential intermediate gets ``lucky'' and quickly forms an
intermediate, or it dissolves back to monomers, leaving few dimers,
trimers, and so on.

\begin{figure}
\includegraphics{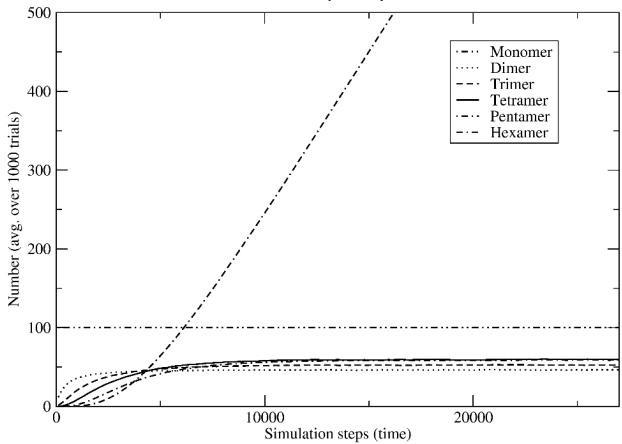}
\caption{Number of each size as a function of time (simulation steps) with nonzero breaking. Compare to figure 6; note that the number of intermediates reaches equilibrium faster and at smaller numbers but the hexamer number still grows linearly at long times.}
\end{figure}

The kinetics equations we can write down to describe this simulation
are relatively simple. With $r_{nm}$ the rate constant for forming
m-mers from n-mers, and $b_{nm}$ the rate of breaking n-mers into
m-mers plus monomers, we can write:
\begin{equation}
[c_1]=c
\end{equation}
 
\begin{equation}
\frac{d[c_2]}{dt}=r_{12}[c_1]^2-r_{23}[c_1][c_2]-b_{21}[c_2]+b_{32}[c_3]
\end{equation}

\begin{equation} 
\frac{d[c_3]}{dt}=r_{23}[c_1][c_2]-r_{34}[c_1][c_3]-b_{32}[c_3]+b_{43}[c_4]
\end{equation}

\begin{equation}
\frac{d[c_4]}{dt}=r_{34}[c_1][c_3]-r_{45}[c_1][c_4]-b_{43}[c_4]+b_{54}[c_5]
\end{equation}

\begin{equation}
\frac{d[c_5]}{dt}=r_{45}[c_1][c_4]-r_{56}[c_1][c_5]-b_{54}[c_5]
\end{equation}

\begin{equation}
\frac{d[c_6]}{dt}=r_{56}[c_1][c_5]
\end{equation}

Since we know that the hexamer number grows linearly at steady state
and all of the other concentrations are unchanging, we can greatly
simplify the above kinetics by looking at the steady state only. We
can work backwards from the steady state behavior of the hexamers to
find the dependence of the steady state rate of hexamer formation on
the different kinetic parameters and ultimately on the monomer
concentration.

This straightforward kinetics analysis produces the equlibrium result
\begin{equation}
m=\frac{r_{12} c^2}{1+\frac{b_{21}}{r_{23} c} \left\{1+\frac{b_{32}}{r_{34} c}\left[1+\frac{b_{43}}{r_{45} c}\left(1+\frac{b_{54}}{r_{56} c}\right)\right]\right\}}
\end{equation} where $m$ is the slope at equilibrium of the hexamer formation rate.

The constants in our simple result for $m$, above, can be measured
from our simulation. However, our simulation does not necessarily
reproduce what these constants would be in a biological system. So it
is difficult to say exactly what the rate of intermediate formation,
$m$, would be in a real system. However, it is nevertheless useful to
know the functional form of its dependence on the monomer
concentration.

The result that the hexamer number begins growing linearly eventually
is independent of monomer concentration. This is important because it
means some hexamers can form given these simple rules even if breaking
dominates. Given that result, it seems safe to assume that if
hexameric intermediates are stable, some will form in biological
systems.

In our model, the hexamer number grows linearly indefinitely, which is
obviously unrealistic biologically. The reason for this is that we
include no mechanism to remove finished hexamers. Realistically, they
would be cleared from the body somehow. They could be endocytosed from
the cell surface and degraded via the proteasome mechanism or some
other pathway. Additionally, any hexamers being taken up into aggregates would reduce this number. Regardless, realistically the number should stabilize
at some fixed value determined by the balance of the clearance rate
and the formation rate.

	
With the result that some hexamers form even at low monomer
concentrations (and more would form if they are trimers), a model was developed where now hexameric
intermediates occupy a single cell on the lattice (equivalently, these could be trimeric intermediates). This model,
described below, largely maintains the same attractive features of the
original, showing that if areal aggregation is the explanation for
these features, as we suggested, this aggregation could be of
hexameric intermediates.

Part of our basis for this model is the observation that the
intermediates are not yet stably misfolded since formation of
intermediates in studies of yeast prions does not lead to a change in
circular dichroism results; it is only when they aggregate with a seed
that they stably misfold (Serio et al., 2002). If any aggregates
consist of oligomers like those observed by Wille et al., then any
misfolding of these intermediates must be catalyzed by existing
aggregates or few-hexamer oligomers. We hypothesize that the mechanism
for this is intermediates forming bonds to an existing seed. When
solvent is excluded locally around these oligomers and their neighbors
include a misfolded oligomer or aggregate, they misfold. The important
point is that it is solvent exclusion around an intermediate that can
cause it to misfold, making this a very rare sporadic event. But a
misfolded seed can help this process by providing a place where
intermediates bond, helping the solvent exclusion process. These rules
make this model essentially identical in terms of kinetics to our
original model.

Details of the algorithm for this model and mapping are shown in figure 10.
	
\begin{figure}
\includegraphics{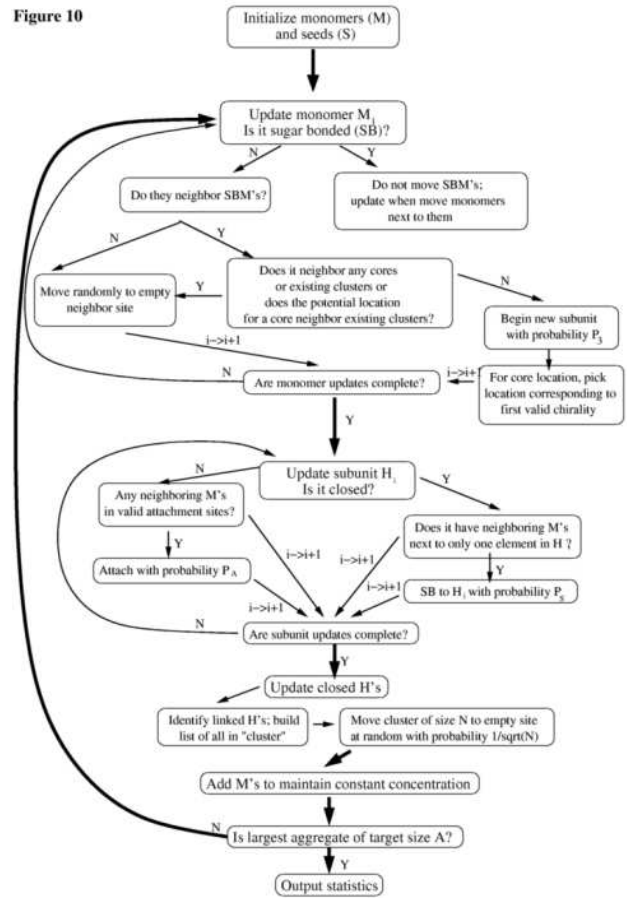}
\caption{Flow chart for simulation mapping back into our original model. Here we basically have free monomers (fM's), attached monomers that are not yet stably misfolded (aM) and monomers that have stably misfolded and aggregated (H). We have some choice of a parameter, $Q_{bc}=m$. This model will capture the features of our original model for m between 3 and 6. We compute $N_{bc}$, the bond coordination number, with $N_{bc}=n_{fM}+n_{aM}+(m-1)*n_{H}$ where the n's are the number of neighboring $fM$'s and so on. $N_{hc}$, the ``hardening'' or aggregating coordination number, is given by $N_{hc}=n_{fM}+n_{aM}+n_{H}$. We refer to $Q_{bc}$ as the bonding critical coordination number and $Q_{hc}$ as the ``hardening'' critical coordination number.}
\end{figure}

However, from our old model we estimated the sporadic form of the
disease could have a peak around 1000 years given a biological
concentration of $10^{-3}$\%. In our new model we find that it is very
difficult to estimate this number as the scaling of the time as a
function of monomer concentration is complicated. It was hoped that
this model would give a result for the onset of sporadic disease that
could be compared with the time for onset of the infectious form to
see if the results were consistent with the roughly 1 in $10^6$
incidence of sporadic CJD that we earlier pointed out. Unfortunately,
it is difficult for our model to give a concrete answer at this time
as the answer depends too much on the value of the biological monomer
concentration. We do find, however, that the power law used previously to scale the sporadic data, $c^{-3}$, is a lower bound on the separation. That is, the actual exponent should be larger, meaning that we previously underestimated the separation of timescales. Thus although we cannot say exactly what the separation of timescales here will be, we can say that it will be greater than the two orders of magnitude that we previously estimated.

This work suggests that a model like our earlier one, modified to
involve areal aggregation of hexameric or trimeric intermediates, could maintain
the same attractive features of our earlier model in explaining
certain aspects of the diseases. However, without precise knowledge of
the biological monomer concentration and a way to measure relevant
rate constants, it is difficult to make numerical predictions from
this model.

\section{DISCUSSION}

Our work has shown that both in the case of monomer addition to a
seed, and in the case of growth via intermediates, it is possible to
produce aggregates like those observed by Wille et al. This leaves the
question of how such aggregates actually grew. If areal aggregation is
the cause, or part of the cause, of the difference between lag and
doubling times, as suggested by Slepoy et al., then our work suggests
that intermediates are already present \emph{in vivo} prior
to aggregation.

Our work has also shown that a model can be developed which, with
suitable parameters, can reproduce areal aggregates like those
actually observed while maintaining the same features of our original
model.

Whether or not areal aggregation is actually important in these
diseases, we can gain insight from this model. If the aggregates
observed are growing via monomer addition, we gain some constraints on
the structure simply from our rules. On the other hand, if
intermediates are important to aggregation, then our results indicate
the intermediate concentration can be quite important. At high
intermediate concentrations, intermediates form relatively
fast. However, at low intermediate concentrations, intermediate
formation time scales as $1/c^6$. This result is exciting because it
suggests the intermediates as a target to prevent aggregation. Simply
reducing the monomer concentration by a factor of 2 would
decrease the number of intermediates by a factor of $2^6$ or
64. Within our model, this would certainly increase the aggregation
time, and thus slow down the disease, by at least the same factor. For
a disease which can typically incubate for years, this obviously would
be a great advantage.

In this case, the location of the transition between low concentration
behavior and high concentration behavior is, roughly speaking, set by
the probability of monomers breaking off from an intermediate before
it becomes a stable hexamer. Thus if this probability could be
increased slightly --- that is, the bonds between monomers could be
weakened slightly --- it would have the result described above. This
could provide an explanation for one experimental observation. Humans
have a methionine/valine polymorphism at codon 129 of the gene for the
prion protein. To date, everyone affected by vCJD has been
methionine/methionine homozygous. This effect was also seen in the
prion disease Kuru, where the methionine/methionine genotype was
associated with increased susceptibility and the shortest incubation
time (Goldfarb, 2002). If replacing methionine with valine weakened
the monomer-monomer bonds within a forming intermediate and reduced
intermediate concentration, this could have exactly the effect
described above.

In all, our work shows that our earlier model can be extended to
produce aggregates like those observed \emph{in vitro} while still
maintaining its attractive features. Our work also suggests possible
mechanisms for formation of these aggregates. If the aggregates form
by monomer addition, it constrains protein structure. If they form by
addition of intermediates, it highlights the importance of bonds
within the intermediates as a target for possible treatment strategies. Our model suggests that an experiment to measure the biological intermediate
concentration, if there is such a concentration, would be very useful. That would indicate whether such intermediates are present at a high enough
concentration to be important biologically. Additionally, this work
suggests that experimentalists should check and see whether
reasonably-sized aggregates of prion protein can be found \emph{in
vivo} on the cell surface. This confinement to the cell surface
concievably could make the difference between the 1D fibrillar
aggregates typically observed \emph{in vitro} and 2D areal aggregates
like those suggested by the model of Slepoy et al. Direct measurements, or detailed simulations, giving the strengths of beta bonds between monomers compared to bonds between subunits would be very useful.

One general scheme for experimentally testing the possible role of intermediates and estimating their concentration is via spin labelling (Hubbell 1998, Columbus 2002).  Briefly, a small molecule with a free spin can preferentially react and attach to cysteine residues in proteins.  Frequently, these residues are moved around a protein via mutagenesis to then map out structures, but for these purposes a less refined approach is required.  Since the PrP protein already possesses cysteine residues at the position of the disulfide bond, the spin labels can attach there (and will not disrupt the disulfide bond).  Then the spin-spin interactions will produce a different characteristic spectrum for monomers, incomplete intermediates and complete intermediates, in particular, with a progressive broadening upon moving from monomers to complete intermediates.  Since the spins can have interactions with other spins within a 3 nm sphere, we don't doubt that the broadening will be observable.  Of course, since the spin labels will react with \emph{any} cysteines present, it is important to carry this out first in \emph{in vitro} aggregation experiments with purified prion extracts.  This will help to identify the conditions which can lead to areal aggregation as observed by Wille et al. (2000), and serve as an existence proof at least for significant oligomeric intermediate concentrations.  
	     
\small	
This research is supported in part by the NEAT-IGERT program sponsored by the National Science Foundation (IGERT Grant DGE-9972741).

We gratefully acknowledge fruitful discussions on spin labelling with John Voss, and on areal prion aggregates with
Holger Wille.   R.R.P.S. and D.L.C. have benefitted from
discussions at workshops of the Institute for Complex Adaptive
Matter.

Sandia is a multiprogram laboratory operated by Sandia Corporation, a
Lockheed Martin Company, for the United States Department of Energy's
National Nuclear Security Administration under contract
DE-AC04-94AL85000.

\newpage

\large
\textbf{References: }

\normalsize
Bruce, M.E., Will, R.G., Ironside, J.W., McConnell, I., Drummond, D., Suttie, A., McCardle, L., Chree, A., Hope, J. Birkett, C., Cousens, S., Frasier, H., Bostock, C.J.   1997. Transmissions to mice indicate that 'new variant' CJD is caused by the BSE agent. \emph{Nature} 389: 498-501.

Caughey, B. 2000. Prion protein interconversions. \emph{Philos. Trans. R. Soc. Lond. B Biol. Sci.} 356:197-202.

Columbus, L. and Hubbell, W.L. 2002. A new spin on protein dynamics. \emph{Trends Biochem. Sci.} 27: 288-295. 

Come, J.H., Fraser, P.E., Lansbury, P.T. Jr. 1993. A kinetic model for amyloid formation in the prion diseases: Importance of seeding. \emph{Proc. Nat. Acad. Sci. USA} 90: 5959-5963. 
 
Goldfarb, L.G. 2002. Kuru: The old epidemic in a new mirror. \emph{Microbes Infect.} 4: 875-882.

Hubbell, W.L., Gross, A., Langen, R., and Lietzow, M.A. 1998.  Recent advances in site-directed spin labelling of proteins.  \emph{Curr. Opin. Struct. Biol.}, 8:  649-656.

Hill, A.F., Desbruslais, M., Joiner, S., Sidle, K.C.L., Gowland, I., Collinge, J., Doey, L.J., Lantos, P. 1997. The same prion strain causes vCJD and BSE. \emph{Nature} 389: 448-450.

Kulkarni, R., Slepoy, A., Singh, R.R.P., Pazmandi, F. 2003. Theoretical modeling of prion disease incubation. \emph{Biophys. J.} In press.

Masel, J., Jansen, V.A.A., Nowak, M.A. 1999. Quantifying the kinetic parameters of prion replication.  \emph{Biophys. Chem.} 77: 139-152.

Philips, Lord of Worth Malraven, Bridgeman, J., Ferguson-Smith, M. October 2000. \emph{Report of the BSE Inquiry}, Volume 1, Chapter 12, Paragraph 1122(i). http://www.bseinquiry.gov. Jan. 4, 2003.

Safar, J.G., Scott, M., Monaghan, J., Deering, C., Didorenko, S., Vergara, J., Ball, H., Legname, G., Leclerc, E., Solforosi, L., Serban, H., Groth, D., Burton, D.R., Prusiner, S.B., Williamson, R.A. 2002. Measuring prions causing bovine spongiform encephalopathy or chronic wasting disease by immunoassays and transgenic mice. \emph{Nature Biotech.} 20: 1147-1150.

Scott, M.R.D., Telling, G.C., Prusiner, S.B. 1996. Transgenetics and Gene Targeting in Studies of Prion Diseases. \emph{In} Prions Prions Prions, S.B. Prusiner, editor. Springer-Verlag, Berlin, Heidelberg. 95-123. 

Scott, M.R., Will, R., Nguyen, H.-O. B., Tremblay, P., DeArmond, S., Prusiner, S.B. 1999. Compelling transgenetic evidence for transmission of bovine spongiform encephalopathy prions to humans. \emph{Proc. Nat. Acad. Sci. USA} 96(26):15137-15142

Slepoy, A., Singh, R.R.P, P\'{a}zm\'{a}ndi, F., Kulkarni, R.V., Cox, D.L. 2001. Statistical Mechanics of Prion Diseases. \emph{Phys. Rev. Lett.} 87(5): 581011 to 581014.

Serag, A.A., Altenbach, C., Gingery, M., Hubbell, W.L., Yeates, T.O. 2002. Arrangement of subunits and ordering of beta-strands in an amyloid sheet. \emph{Nature Struct. Biol.} 9: 734-739. 

Serio, T.R., Cashikar, A.G., Kowal, A.S., Sawicki, G.J., Moslehi, J.J. Serpell, L., Arnsdorf, M.F., Lindquist, S.L. 2000. Nucleated conformational conversion and the replication of conformational information by a prion determinant. \emph{Science} 289: 1317-1321.

Wille, H., Michelitsch, M.D., Guenebaut, V., Supattapone, S., Serban, A., Cohen, F.E., Agard, D.A., Prusiner, S.B. 2002. Structural studies of the scrapie prion protein by electron crystallography. \emph{Proc. Nat. Acad. Sci. USA} 99(6): 3563-3568.

Weissmann, C., Enari, M., Klohn, P-C., Rossi, D., Flechsig, E. 2002. Transmission of Prions. \emph{J. Infect. Dis.} 186(Suppl. 2): S157-65.

\end{document}